\documentclass[aps,prl,twocolumn,superscriptaddress,nofootinbib]{revtex4-2}
\usepackage{amsmath,amssymb}
\usepackage{graphicx}
\usepackage{tikz}
\usetikzlibrary{patterns}
\usepackage{hyperref}

\begin{document}

\title{The KM3NeT 220 PeV event: a neutrino messenger from the grand unification scale?}

\author{Vernon Barger}
\affiliation{Department of Physics, University of Wisconsin--Madison, Madison, WI 53706, USA}

\date{July 14, 2026}

\begin{abstract}
The 220~PeV neutrino KM3-230213A, in tension with IceCube and Auger unless narrow, admits a two-body reading: relic decay $X \to \nu\bar\nu$ with $m_X \approx 4.4\times10^{8}$~GeV. Consistency with neutrino and $\gamma$-ray bounds sets $\tau/f \approx 10^{29}$--$10^{30}$~s, lifetime over relic fraction. For a flavored $^1S_0$ onium, the operator scale lands in the Majorana-mass range of unified models: one Higgs insertion gives a singlet mediator, $M/|C_7|^{1/3} \approx 10^{14}$--$10^{16}$~GeV across the allowed relic fraction; two give the seesaw window; Dirac or Majorana neutrinos select the branch. We give the construction and its decisive tests.
\end{abstract}

\maketitle

\begin{figure}[t]
\centering
\resizebox{\columnwidth}{!}{%
\begin{tikzpicture}[font=\footnotesize]
\def\dx{1.02}
\def\ymax{5.4}
\draw[->,thick] (-0.1,0) -- (8.7,0);
\draw[->,thick] (0,-0.1) -- (0,\ymax);
\node[below=5pt] at (4.1,-0.35) {$E_\nu$};
\node[rotate=90] at (-0.75,2.7) {$E^2\Phi$ \ (schematic)};
\foreach \e in {13,15,17.34,18,21}{
  \pgfmathsetmacro\xx{(\e-13)*\dx}
  \draw (\xx,0) -- (\xx,-0.12);
}
\foreach \e/\lab in {13/{10\,TeV},15/{PeV},18/{EeV},21/{ZeV}}{
  \pgfmathsetmacro\xx{(\e-13)*\dx}
  \node[below] at (\xx,-0.12) {\lab};
}
\pgfmathsetmacro\xpk{(17.34-13)*\dx}
\pgfmathsetmacro\xgapL{(16.2-13)*\dx}
\pgfmathsetmacro\xgapR{(17.9-13)*\dx}
\fill[black!6] (\xgapL,0) rectangle (\xgapR,\ymax);
\node[align=center,text width=1.7cm] at (3.58,3.10)
  {\scriptsize integral limits\\ target other\\ energies};
\pgfmathsetmacro\xicL{(13-13)*\dx}
\pgfmathsetmacro\xicR{(16.2-13)*\dx}
\def\yic{2.3}
\draw[thick] (\xicL,\yic) -- (\xicR,\yic);
\fill[pattern=north east lines,pattern color=black!45,opacity=0.5]
  (\xicL,\yic) rectangle (\xicR,\ymax);
\node[above,align=center] at ({(\xicL+\xicR)/2},\yic)
  {\scriptsize IceCube\\ \scriptsize (no tail here)};
\pgfmathsetmacro\xauL{(17.9-13)*\dx}
\pgfmathsetmacro\xauR{(21-13)*\dx}
\def\yau{2.75}
\draw[thick] (\xauL,\yau) -- (\xauR,\yau);
\fill[pattern=north east lines,pattern color=black!45,opacity=0.5]
  (\xauL,\yau) rectangle (\xauR,\ymax);
\node[above,align=center] at ({(\xauL+\xauR)/2},\yau)
  {\scriptsize Auger, ARA/RNO-G\\ \scriptsize (nothing above)};
\def\yieh{4.0}
\draw[thin,black!55] (\xgapL,\yieh) -- (\xgapR,\yieh);
\fill[pattern=north east lines,pattern color=black!20,opacity=0.4]
  (\xgapL,\yieh) rectangle (\xgapR,\ymax);
\node[above,align=center] at (4.05,\yieh)
  {\scriptsize IceCube EHE\\ \scriptsize differential};
\fill (\xpk,\yieh) circle (0.045);
\node[below right,inner sep=1.5pt] at (\xpk,\yieh) {\scriptsize $\sim\!2\sigma$};
\def\pkH{2.0}
\draw[very thick]
  plot[smooth,domain=-3.4:3.4,samples=60]
  (\xpk + \x*0.30, {\pkH*exp(-\x*\x)});
\draw[dashed] (\xpk,0) -- (\xpk,\pkH);
\node[above,align=center] at (\xpk,\pkH) {\scriptsize 220\,PeV feature};
\end{tikzpicture}%
}
\caption{Schematic of the spectral argument. A narrow feature at 220~PeV,
whether the relic line developed here or the resonant astrophysical peak,
falls in the energy window (shaded) between the IceCube TeV--PeV band
and the EeV-band searches of Auger and the in-ice radio
arrays~\cite{ARA:2022,ARA:2026}. Dark hatched regions indicate where those
searches are most sensitive; they are flux-, spectrum-, flavor-, direction-, and
duration-dependent and must be recast for the assumed line rather than read
as hard exclusions in energy, and electroweak radiation adds a continuum
below the line. The lighter band marks IceCube's differential
extremely-high-energy sensitivity, which does span the window; the dot marks
the residual $\approx 2\sigma$ exposure tension at the line energy quantified
in the text~\cite{IceCube:EHE,KM3NeT:landscape}. The heuristic point is only
that a feature narrow in energy is weakly constrained by searches targeting
other energies.
Levels are indicative, not a fit.}
\label{fig:window}
\end{figure}
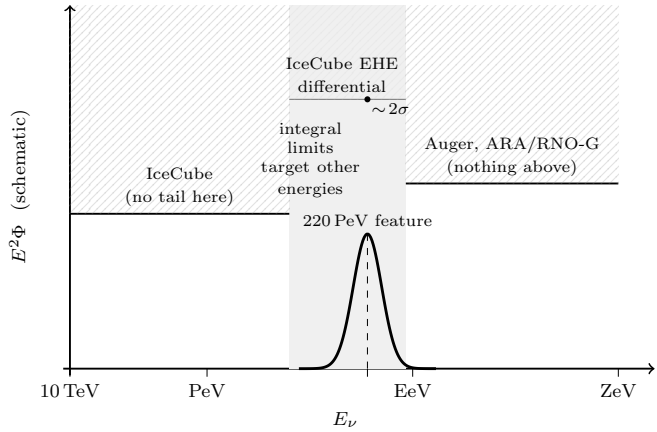

The KM3NeT/ARCA detection of the muon event KM3-230213A, with reconstructed neutrino energy near 220~PeV and a 90\% interval from 72~PeV to 2.6~EeV, is the highest-energy neutrino yet reported~\cite{KM3NeT:Nature}. Its interpretation is set as much by non-detections as by the event. IceCube reports no comparable event~\cite{IceCube:EHE}, the Pierre Auger Observatory none in the EeV band~\cite{Auger:2019,Auger:PS2019,Auger:ICRC2023}, and Baikal-GVD sets consistent multi-PeV constraints~\cite{Baikal:2025}; a joint analysis finds a mild to moderate $1.6\sigma$--$2.9\sigma$ tension for a diffuse isotropic flux, mildest for a transient point source~\cite{KM3NeT:landscape}, and a cosmogenic origin is disfavored by the same EeV-band limits~\cite{KM3NeT:cosmogenic}. A narrow spectral feature at the event energy relaxes the cross-energy constraints once each is recast for a line of the assumed flux, flavor, direction, and duration (Fig.~\ref{fig:window}): a spectrum concentrated near 220~PeV is weakly constrained by the IceCube TeV--PeV limits, and the EeV-band limits from Auger and the in-ice radio arrays~\cite{ARA:2022,ARA:2026} are most sensitive above the line energy, pending recasts with their full response. What no spectral shape removes is the residual $\sim 2\sigma$ at the peak energy itself, an exposure effect carried by transience or by rarity, not by shape~\cite{KM3NeT:landscape,Neronov:2025}.

Two mechanisms make a narrow feature, and they are not equally narrow. An astrophysical accelerator can produce a resonant peak, but pion-decay kinematics typically broaden it to a factor of $\sim 2$--$3$ in energy for a quasi-monoenergetic beam, the exact width depending on the parent spectrum, interaction angles, and detector response. A line, narrow to detector resolution, has a particularly economical origin, two-body decay $X \to \nu\bar\nu$ of a relic with $m_X = 2E_\nu \approx 4.4\times10^{8}$~GeV, a reading with an active literature~\cite{Kohri:2025,Borah:2025}. This Letter develops the line reading into a quantitative, explicitly conditional claim: \emph{if} the feature is a two-body decay line from a cosmologically long-lived flavored onium, and \emph{if} the leading decay operator carries one electroweak-Higgs insertion with an $O(1)$ coefficient, then the neutrino energy sets the relic mass, the flux sets the lifetime-to-fraction ratio, and the lifetime sets the scale of the symmetry breaking that lets the relic decay, landing in the Majorana-mass window of unified models. We state the chain, link by link, the astrophysical alternative, and the tests that decide.

\textit{The messenger chain.} Three measured or inferred numbers map onto three theory scales. (i) Under the Galactic-endpoint hypothesis the energy fixes the mass, $m_X = 2E_\nu(1+z_{\rm origin}) \to 2E_\nu \approx 4.4\times10^{8}$~GeV for local decay, with the 90\% interval spanning $1.4\times10^{8}$ to $5\times10^{9}$~GeV; the reconstructed energy assumes a spectral prior (End Matter); one event cannot tell endpoint from shoulder. (ii) The flux fixes the combination $f/\tau$, with $f = \Omega_X/\Omega_{\rm DM}$ the fraction of the dark matter residing in $X$ ($f = 1$: $X$ is all of it) and $\tau$ its lifetime; the inferred value is analysis-dependent. A collaboration fit treating the event as typical returns $\tau \approx 10^{26}$--$10^{27}$~s, in tension with neutrino and $\gamma$-ray bounds~\cite{KM3NeT:hdm}; consistency with those bounds gives $\tau/f \approx 10^{29}$--$10^{30}$~s~\cite{Kohri:2025,Borah:2025,Barman:2025,Aloisio:2025}. A public-exposure estimate anchored to the joint flux reproduces $\tau/f \approx (1$--$3)\times10^{29}$~s; the collaboration's detector-level $\nu\bar\nu$ fit, making the event typical of KM3NeT alone, returns the shorter value (End Matter). The ranges differ by $10^{4}$ in $\tau$ but a factor $\sim 5$ in the scale. (iii) The lifetime, through the bound-state decay rate below, fixes an effective four-fermion coupling $G_{\rm eff}$, and the operator that generates $G_{\rm eff}$ converts it into a mediator mass once the symmetry-breaking insertions are identified. The chain terminates, in its minimal form ($\tau/f = 10^{30}$~s, $C_7 = 1$), at $M \approx 10^{16} f^{1/6}$~GeV: $5\times10^{15}$~GeV at the fiducial $f = 10^{-2}$ and $10^{16}$~GeV at $f = 1$; the broad-channel best-fit lifetime would give $2$--$3\times10^{15}$~GeV but does not describe the line (End Matter). The flux fixes only $f/\tau$: both the Galactic and the extragalactic decay fluxes scale as $f/\tau$, so the sky gradient and redshift structure (below) test the decay hypothesis and constrain the halo model but do not separate $f$ from $\tau$. The microscopic lifetime, and hence the scale, therefore carry an explicit relic-fraction dependence, $\tau = f\,(\tau/f)$ and $M \propto \tau^{1/6} \propto f^{1/6}$ at fixed $f/\tau$; the numbers below take the fiducial $f = 10^{-2}$, motivated by an axion-dominated dark sector (End Matter); $f = 1$ bounds the scale.

\textit{Why an onium.} A two-fermion final state $\nu\bar\nu$ requires a bosonic parent; among bosonic options we take a flavored $^1S_0$ onium $\bar Q' Q$, the case that ties the lifetime to a high-scale operator. A single fermion's two-body decay is instead $\nu h$ or $\nu Z$; an elementary boson is viable but has no bound-state relation between rate and scale. A same-flavor onium annihilates promptly through its own gauge interactions ($\Gamma \sim \alpha_{\rm eff}^5 m_X$) and cannot survive cosmologically; the relic must be the flavored combination, stable under an accidental flavor symmetry of the dark sector and decaying only through the operator that breaks it. Stability places a sharp requirement on that symmetry. Quantum gravity is expected to violate global symmetries, and a Planck-suppressed dimension-6 operator $(\bar Q' i\gamma_5 Q)(\bar\nu\nu)/M_{\rm Pl}^2$ would give $\tau \sim 10^{12}$--$10^{15}$~s (reduced to unreduced $M_{\rm Pl}$), short of the age of the universe by several orders of magnitude regardless of convention; the flavor symmetry therefore cannot be an accident of the low-energy Lagrangian alone but must be protected, most economically as a discrete gauge symmetry, which survives gravitational violation~\cite{Krauss:1989}. The selected operator then preserves the gauged $Z_N$ and breaks only the accidental continuous flavor symmetry, which requires a lepton-sector field to carry a compensating dark charge; a $\nu_R$ so charged has its Dirac Yukawa forbidden by the same symmetry, so the $\nu_R$ entering the decay operator must be a light sterile state distinct from those generating the Dirac masses. The protection requirement is the reason the decay proceeds only through the high-scale operator whose coefficient the lifetime then measures. The bosonic channel is also the $\gamma$-quiet one: a fermionic relic decaying through $\nu h$ puts half the energy into a Higgs cascade and faces stronger $\gamma$-ray constraints, while $\nu\bar\nu$ deposits everything into neutrinos up to electroweak radiation. The symmetry can be as small as a $Z_N$ under which $Q$ and $Q'$ carry different charges, provided the assignment is anomaly-free, stabilizes the flavored onium, forbids every operator below dimension seven, admits the selected operator, and induces no faster channel through the mediator sector; the explicit charge table is completion-level work not performed here. The charge assignment fixes which insertion, and hence which operator dimension, first connects the dark bilinear to neutrinos.

\textit{The lifetime as a scale meter.} The decay is a bound-state annihilation, so the rate factorizes into the wave function at contact and the constituent annihilation through the effective operator,
\begin{align}
\Gamma_d(X \to \nu\bar\nu) &= |\psi(0)|^2\,\langle\sigma v\rangle_d
\equiv \kappa\, \alpha_{\rm eff}^3\, m_X^5\, G_{\rm eff}^2,
\label{eq:onium} \\
G_{\rm eff} &\sim \frac{v_H^{\,d-6}}{M^{\,d-4}}, \nonumber
\end{align}
with $|\psi(0)|^2 = (\alpha_{\rm eff}\, m_X/4)^3/\pi$ for the Coulombic $^1S_0$ ground state, the state a pseudoscalar dark bilinear couples at leading order (a scalar bilinear instead projects onto the velocity-suppressed $P$ wave), $\langle\sigma v\rangle_d \sim G_{\rm eff}^2\, m_X^2$, $\alpha_{\rm eff} \approx 0.05$, and $\tau_d = \Gamma_d^{-1}$; the factor $\kappa$ collects the wave-function normalization, spin projection, phase space, dark-color factors, and active-neutrino multiplicity, equal to $1/(64\pi)$ for the displayed factors alone, with the complete spin-projected calculation for a specified dark gauge group modifying it at $O(1)$; equal constituent masses $m_Q = m_X/2$ are assumed throughout. Here $M$ is the mediator scale and $v_H$ the vev of the field supplying the $(d-6)$ flavor-breaking insertions of a dimension-$d$ operator. The benchmark $\alpha_{\rm eff} \approx 0.05$ is the dark coupling evaluated at the Bohr momentum $\alpha_{\rm eff} m_X/4$; any value compatible with confinement below that momentum serves, and the inference inherits only the mild $M \propto \alpha_{\rm eff}^{1/2}$ sensitivity noted below. The observable fixes the coupling, not the scales separately: $\tau \approx 10^{30}$~s at $m_X = 4.4\times10^{8}$~GeV gives
\begin{equation}
G_{\rm eff} \approx 1.8\times10^{-47}\,\kappa^{-1/2}~\mathrm{GeV}^{-2},
\end{equation}
i.e., $2.5\times10^{-46}~\mathrm{GeV}^{-2} \approx 2\times10^{-41}\,G_F$ at the Coulombic normalization.
The decomposition then reads off the ultraviolet. A pure four-fermion operator ($d=6$, no insertion) needs $M \approx 0.6$--$2\times10^{23}$~GeV, trans-Planckian, and is forbidden by the gauged flavor symmetry above. The leading allowed operator carries one insertion. If that insertion is the electroweak Higgs, e.g., $\mathcal{O}_7 = (\bar Q' i\gamma_5 Q)(\bar L \tilde H \nu_R)/M^3$ for Dirac neutrinos (each decay pairs an active neutrino with its sterile-helicity partner, an $O(1)$ factor folded into $f/\tau$), built from the gauge-invariant Dirac-neutrino Yukawa structure $\bar L \tilde H \nu_R$ (for systematic operator bases with right-handed neutrinos see~\cite{Lehman:2014,Bhattacharya:2015,Liao:2017}), then $v_H = v \approx 246$~GeV and the operator coefficient $C_7$ enters through $G_{\rm eff} = C_7\, v/M^3$, so the observable fixes $M/|C_7|^{1/3}$, not $M$ alone:
\begin{align}
\frac{M}{|C_7|^{1/3}} &\simeq 1.0\times10^{16}~\mathrm{GeV}\left(64\pi\kappa\right)^{1/6}
\left(\frac{\alpha_{\rm eff}}{0.05}\right)^{1/2} \nonumber \\
&\quad\times
\left(\frac{\tau}{10^{30}~\mathrm{s}}\right)^{1/6}
\left(\frac{m_X}{4.4\times10^{8}~\mathrm{GeV}}\right)^{5/6}.
\label{eq:gut}
\end{align}
At the Coulombic $\kappa = 1/(64\pi)$, with $\tau = f(\tau/f)$ at $\tau/f = 10^{30}$~s, this is $M/|C_7|^{1/3} \approx 10^{16} f^{1/6}$~GeV, i.e., $5\times10^{15}$~GeV at the fiducial and $10^{16}$~GeV at $f = 1$; every rate-normalization uncertainty enters through the sixth root. Two insertions give the onium-dressed Weinberg structure~\cite{Weinberg:1979} $\mathcal{O}_8 = (\bar Q' i\gamma_5 Q)(LH)(LH)/M^4$, a $\Delta L = 2$ operator producing $\nu\nu$ at the same line energy, with $M \approx 2$--$8\times10^{12}$~GeV across $\kappa$ and $f = 10^{-2}$--$1$, the classic seesaw window~\cite{Minkowski:1977,Yanagida:1979,GellMann:1979,Mohapatra:1980}. If the insertion is instead a high-scale flavon with $\langle\Phi\rangle \sim m_X$, the momentum-scaling estimate applies and $M$ returns toward $10^{18}$~GeV. The minimal single-Higgs reading is the one we headline; it carries a structural link, in that the required insertion is the same $LH$ combination that enters the neutrino-mass operator, connecting the relic lifetime, the inferred scale, and the neutrino-mass sector within this reading. The fork between the two readings is itself a neutrino-physics statement. The single-insertion operator $\mathcal{O}_7$ requires a light right-handed neutrino, natural alongside Dirac-type neutrino masses; if neutrinos are Majorana and no light $\nu_R$ exists, the leading operator is $\mathcal{O}_8$ and the meter reads the seesaw scale instead. Either way the line is a messenger from the far-ultraviolet neutrino sector; under the minimal assumption that the leading allowed operator dominates, which decade it reports is tied to the Dirac-versus-Majorana question. The correlation is a feature of the minimal reading, not of generic models, and it breaks if $\mathcal{O}_7$ and $\mathcal{O}_8$ contribute comparably or if a flavon supplies the insertion.

The scale is stable precisely because of the sixth root: the full lifetime range $10^{26}$--$10^{30}$~s spans $M/|C_7|^{1/3} \approx 2\times10^{15}$--$10^{16}$~GeV, and the 90\% energy interval, entering through $G_{\rm eff} \propto m_X^{-5/2}$, together with the span of $\kappa$ and a relic fraction anywhere down to the survival floor $f \gtrsim 4\times10^{-12}$ (below which the population has decayed), widens this to roughly $10^{14}$--$10^{17}$~GeV. The bound-consistent normalization places the mediator within a factor of two of the unification scale, $\approx 5\times10^{15}$~GeV at the fiducial $f = 10^{-2}$ and $10^{16}$~GeV at $f = 1$; the collaboration best fit, in tension with $\gamma$-ray bounds within the decay hypothesis, would lower it to $\approx 2\times10^{15}$~GeV. We headline the bound-consistent value. The constituents themselves are Standard Model singlets charged under a dark color group (see below), so gauge coupling unification is untouched, and the mediator at $M$ can belong to the unified sector that communicates the flavor breaking to $\bar L \tilde H \nu_R$; we present this as consistency, not derivation.

\textit{The Majorana window.} The landing points bracket the scale unified models assign to neutrino mass. The seesaw relation $m_\nu = y^2 v^2/M_R$ puts $M_R \approx 6\times10^{14}\, y^2\,(0.05~\mathrm{eV}/m_\nu)$~GeV, so the one-insertion mediator at $(3$--$5)\times10^{15}$~GeV corresponds to $y \approx 2$--$3$ or $m_\nu \approx 0.01$~eV (the former at the edge of perturbativity, $y^2/4\pi \approx 0.3$--$0.7$, favoring the latter reading), and the two-insertion window $10^{12}$--$10^{13}$~GeV to $y \approx 0.05$--$0.1$: both branches sit inside the Majorana-mass range of unified models. The gauge-unification point, $2\times10^{16}$~GeV, probed by proton decay, lies above the landing; the mediator is more naturally the seesaw-sector singlet than the gauge sector, the singlet constituents leaving gauge-coupling running untouched. On the Dirac branch the singlet's Majorana character is a property of the completion, so the statement is an overlap, not a derivation: the selected dimension-seven operator with an $O(1)$ coefficient lands the scale in the range unified models assign to their heavy singlets; we do not construct the completion.

\textit{Production, binding, survival.} The required abundance is minute, which makes production easy. For the benchmark lifetimes, $\Omega_X h^2 = 0.12\,f$ corresponds to a comoving yield $n_X/s \approx 10^{-18} f$. Such a yield is a generic outcome of nonthermal production, not a prediction of a specific reheating temperature: gravitational particle production alone can reach it for suitable inflationary scales at $m_Q \sim 10^{8}$~GeV~\cite{Garny:2016}, and production from the bath during reheating supplies it for $T_{\rm RH}$ well below $m_Q$~\cite{Chung:1998}, the required value depending on the production operator, coupling, and thermal history rather than on the Boltzmann factor alone. A viable completion must supply one such mechanism; we do not fix it here. Keeping $T_{\rm RH}$ below $m_Q$ also avoids a thermalized heavy-quark population, which at $m_Q \approx 2\times10^{8}$~GeV would violate the unitarity bound on thermal relics~\cite{Griest:1990}. Binding requires a Coulombic hierarchy, the inverse Bohr radius $\alpha_{\rm eff} m_X/4$ well above the dark confinement scale; the flavored-onium survival fraction after confinement then depends on the relative $Q,Q'$ abundances, any flavor asymmetry, and the full hadron inventory of the confining sector (same-flavor onia, dark baryons, glueballs, hybrids) with its rearrangement rates, and warrants a dedicated Boltzmann treatment we defer. We take it as an $O(1)$ factor absorbed into $f$, a benchmark assumption rather than a demonstrated result. Dark confinement also disarms the constraint that kills the colored version: constituents with Standard Model color would leave free-quark relics bound into anomalously heavy isotopes, excluded far below any plausible leftover fraction, so the constituents are singlets, vector-like under the dark group.

\begin{figure}[t]
\centering
\resizebox{\columnwidth}{!}{%
\begin{tikzpicture}[font=\footnotesize,xscale=7.0,yscale=1.05]
\draw[->,thick] (0.16,0) -- (1.17,0);
\draw[->,thick] (0.18,-0.02) -- (0.18,3.05);
\node[rotate=90] at (0.115,1.5) {$E^2\Phi$ \ (schematic)};
\node[below] at (0.72,-0.07) {$E_\nu$ (at Earth)};
\foreach \x/\lab in {0.25/{55},0.5/{110},1.0/{220~PeV}}{
  \draw (\x,0) -- (\x,-0.07);
  \node[below] at (\x,-0.07) {\lab};
}
\node at (0.492,0.52) {\scriptsize $(z=1)$};
\fill[black!10]
  plot[smooth,domain=0.2:1.0,samples=120] (\x,{\x/sqrt(0.3/\x^3+0.7)})
  -- (1.0,0) -- (0.2,0) -- cycle;
\draw[thin]
  plot[smooth,domain=0.2:1.0,samples=120] (\x,{\x/sqrt(0.3/\x^3+0.7)});
\draw[very thick]
  plot[smooth,domain=0.2:1.12,samples=400]
  (\x,{(\x<=1)*(\x/sqrt(0.3/\x^3+0.7)) + 2.0*exp(-((\x-1)/0.030)^2)});
\draw[dashed] (1.0,0) -- (1.0,3.0);
\node[above] at (1.0,3.0) {\scriptsize $m_X/2$};
\node[align=center] at (0.47,2.35) {\scriptsize Galactic edge\\ \scriptsize (resolution-narrow)};
\draw[->,thin] (0.62,2.35) -- (0.965,2.30);
\node[align=center] at (0.46,1.66) {\scriptsize extragalactic, redshifted:\\ \scriptsize $E=(m_X/2)/(1+z)$};
\draw[->,thin] (0.36,1.52) -- (0.41,0.21);
\draw[dashed,thin,black!60] (0.449,0) -- (0.449,1.42);
\draw[dashed,thin,black!60] (0.535,0) -- (0.535,1.42);
\node[align=center] at (0.745,1.18) {\scriptsize counterparts:\\ \scriptsize $z=0.87,\ 1.23$};
\draw[->,thin] (0.650,1.24) -- (0.548,1.36);
\draw[->,thin] (0.650,1.12) -- (0.462,1.32);
\node[rotate=90,align=center] at (1.10,1.5) {\scriptsize no flux above $m_X/2$};
\end{tikzpicture}%
}
\caption{Predicted shape of the relic line (schematic). The Galactic
component of the decay flux is an edge at $E = m_X/2$, narrow to detector
resolution; the extragalactic component of the same decay arrives redshifted,
$E = (m_X/2)/(1+z)$, and fills the redshifted shoulder below the edge (shaded),
with $E^2\Phi \propto [(1+z)\sqrt{\Omega_m(1+z)^3+\Omega_\Lambda}\,]^{-1}$
falling to $\approx 30\%$ of its edge value by $z=1$ and $\approx 6\%$ by
$z=3$; the majority of shoulder events lie within a factor of two of the edge.
Nothing lies above $m_X/2$ beyond resolution smearing. Electroweak final-state radiation adds a soft continuum
well below the peak (not shown). Dashed verticals mark $(m_X/2)/(1+z)$ for
the measured counterpart redshifts, $z = 0.870$ (PKS~0605-085) and $z = 1.227$
(PMN~J0606-0724)~\cite{KM3NeT:blazars}: under the line reading, decays at the
counterpart distances populate the shoulder at 118 and 99~PeV, not the
220~PeV edge. Levels are indicative, not a fit.}
\label{fig:lineshape}
\end{figure}
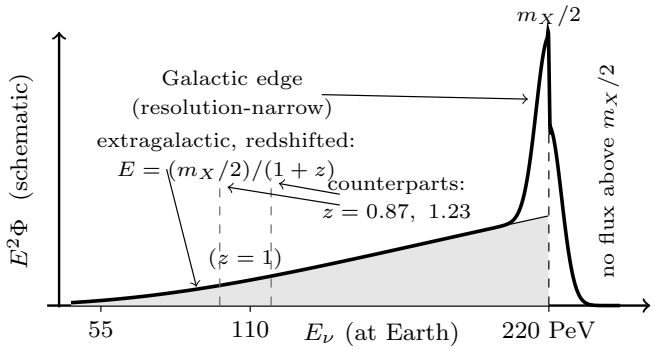

\textit{Line shape and flavor at Earth.} The spike at $m_X/2$ is broadened only by detector resolution and by the electroweak final-state radiation that necessarily dresses a $\nu\bar\nu$ final state at $m_X \sim 4\times10^{8}$~GeV, a soft continuum treated in the relic literature and constrained by $\gamma$-ray observations~\cite{Kohri:2025,Borah:2025}. The full spectral shape carries a signature of its own: the Galactic component is a sharp edge at $m_X/2$, while the extragalactic component of the same decay arrives redshifted, filling a tail below the edge with weight concentrated within a factor of $\sim 2$ ($z \lesssim 1$); the predicted feature is an edge with a redshifted shoulder, not a symmetric bump (Fig.~\ref{fig:lineshape}, End Matter), and the shoulder is far too short to engage the IceCube TeV--PeV limits. Flavor is observed after propagation: over the halo and extragalactic baselines the oscillation phases fully decohere, matter effects are negligible along halo and intergalactic paths even at 220~PeV, and decohered vacuum mixing maps source flavor onto $\sum_i |U_{\alpha i}|^2 |U_{\beta i}|^2$ at Earth. A democratic or incoherent source gives every flavor above $\approx 15\%$; a single mass eigenstate is the exception (pure $\nu_3$: electron fraction of a few percent), so the flavor prediction is operator-dependent, and we quote it for the assumed $\bar L \tilde H \nu_R$ structure with democratic lepton-flavor coefficients, giving a nonzero $\nu_\tau$ component. In the Dirac operator one of the two produced states is the sterile-helicity partner, undetected; the detectable flux is the active member, and for a charge-symmetric $X$--$\bar X$ population it is $\nu/\bar\nu$-symmetric, unlike the $\bar\nu_e$-poor photohadronic alternative below, though $\nu/\bar\nu$ discrimination is statistical at best.

\textit{Sky morphology and rate.} The source is the dark-matter halo, so the flux traces the line-of-sight column, diffuse with no counterpart, no time clustering, and no preferred distance. Because decay integrates the density rather than its square, the anisotropy is gentle; the Galactic-center-to-anticenter ratio is a factor of a few for standard halo profiles, the extragalactic component is isotropic, so the morphology test needs several events. The event direction (RA $94.3^\circ$, Dec $-7.8^\circ$) lies toward the Galactic anticenter, near the halo minimum, where the column falls below the all-sky average by a factor of order two for standard profiles; this is a mild prior against the halo origin. The same $f/\tau$ that yields one event in the partial-array exposure predicts a rate scaling directly with the completed ARCA exposure and with IceCube-Gen2, with arrival directions accumulating toward the Galactic center. The prediction is bounded rather than open-ended: IceCube's larger integrated exposure has not accumulated events at these energies, so the line hypothesis requires the KM3NeT detection to be an upward fluctuation at the level quantified by the joint-exposure analysis~\cite{KM3NeT:landscape}, which also caps the rate expected at the completed arrays; the accompanying electroweak-radiation photons, as computed in the published $\nu\bar\nu$-channel analyses~\cite{Kohri:2025,Borah:2025,Aloisio:2025}, sit near existing bounds; the shower's $O(1)$ chirality and flavor dependence is not recomputed here.

\textit{The astrophysical alternative.} The reading the line must beat is a resonant peak from a flaring accelerator, and it is a serious competitor because candidate counterparts exist. A quasi-monoenergetic proton beam from electrostatic gap acceleration in a black-hole magnetosphere~\cite{BZ:1977,Levinson:2000,Ptitsyna:2016}, at $E_p \approx 20\,E_\nu \approx 4.4$~EeV for fiducial parameters~\cite{Boldt:1999}, interacting at the $\Delta(1232)$ resonance with the thermal infrared field of warm dust, produces a peak (the mechanism of the 30~TeV diffuse break on keV targets~\cite{Winter:2026,KhateeZathul:2026,Barger:APSOS2026}, there with multi-pion production) whose energy measures the target through $\varepsilon_0 E_p \approx 0.16~\mathrm{GeV}^2$, selecting dust at $T \approx 150$--$300$~K; an accretion-flare interpretation with a dust-echo target has been developed for the candidate MRC 0614-083~\cite{Yuan:2025}, and the counterpart study finds a radio flare of PMN~J0606-0724 coincident with the arrival time (pre-trial $0.26\%$, not corrected for trials) and a $\gamma$-ray flare of PKS~0605-085 peaking before it~\cite{KM3NeT:blazars,Dzhatdoev:2025}. The peak is not a line: pion-decay kinematics broaden it even for a delta-function beam, its composition is $\bar\nu_e$-poor and possibly muon-damped~\cite{Kashti:2005}, its parent protons are strongly deflected or decay before arrival, and its cascade photons are delayed and diluted by the intergalactic field. The quantitative backbone, the curvature-radiation ceiling, the photopion optical depth, and the muon-damping condition, is given in End Matter; the point here is the fork. A decisive comparison requires folding both hypotheses through the same KM3NeT energy and angular response, which we do not attempt; at the Letter level, width, sky distribution, and counterpart behavior are the discriminating observables.

\textit{Relation to other new-physics readings.} The event has drawn a range of beyond-Standard-Model interpretations, from modified neutrino propagation and dark-matter-induced effects along the flight path~\cite{Dev:2025} and source-side dark-matter annihilation~\cite{Basu:2025} to general assessments of what new physics the single event can support~\cite{Brdar:2025}. Most such readings modify how an astrophysical neutrino travels or interacts; the line reading is different in kind because it replaces the source, making the feature's width, position, flavor, and sky distribution Lagrangian parameters rather than source properties. The EeV-scale ANITA anomalies~\cite{ANITA:2016,ANITA:2018,ANITAIV:2021}, which have drawn subsurface-reflection and other instrumental readings~\cite{Shoemaker:2020} alongside new-physics ones, lie well above $m_X/2$ and are unconnected to the line, whose flux ends at 220~PeV.

\textit{Spectroscopy of the dark sector.} If events accumulate, the edge position measures $m_X$, the normalization $f/\tau$, the edge-to-shoulder ratio the Galactic-to-extragalactic split and halo column, and the flavor ratios the operator's chirality structure through the averaged mixing (End Matter). Galactic and extragalactic columns are comparable for standard profiles, so edge and shoulder share events roughly evenly, and the split is measured to $\sim N^{-1/2}$: tens of events begin to constrain the split, and of order one hundred approach ten-percent statistical precision, before backgrounds and energy resolution are folded in; the edge/shoulder decomposition is an estimate, not a likelihood forecast.

\textit{What a unified completion must deliver.} The construction is a checklist rather than a model: (i) a vector-like pair $Q, Q'$ of Standard Model singlets charged under a dark color group confining below the Bohr momentum; (ii) a gauged discrete flavor symmetry distinguishing them, forbidding the dimension-6 operator, with the lepton-sector charges compatible with the neutrino Yukawas; (iii) a mediator in the inferred $10^{15}$--$10^{16}$~GeV window coupling the dark bilinear to $\bar L \tilde H \nu_R$, or the double-insertion Weinberg analogue at the seesaw scale; (iv) a nonthermal production mechanism yielding $n_X/s \approx 10^{-18}f$. Each ingredient is individually standard; establishing that they coexist in one unified group, with the onium surviving confinement and no faster decay allowed, is the follow-up work on which a firm claim would depend.

\textit{Tests.} The reading is falsifiable on a short horizon. (i) \emph{Morphology.} Subsequent events at 100--300~PeV tracing the halo column, with a soft gradient toward the Galactic center and no repeating position, support the line; clustering at RA $94.3^\circ$, Dec $-7.8^\circ$, or a second neutrino coincident with a flare of PMN~J0606-0724 or PKS~0605-085, decides for the astrophysical peak. (ii) \emph{Width.} With several events, a spread consistent with detector resolution supports the line, with the redshifted shoulder supplying an internal consistency check once a few events exist; a factor 2--3 spread supports the peak. Any confirmed event at, say, 10~EeV falsifies both single-feature hypotheses. (iii) \emph{Timing.} An IceCube time-dependent search at the event position restricted to a $\lesssim 2$~yr window around February 2023 tests the transient alternative directly~\cite{Neronov:2025}; a null result there disfavors the tested transient models and windows, not generic astrophysical production. (iv) \emph{Companions.} The electroweak-radiation photon shoulder is a diffuse $\gamma$-ray prediction at a level adjacent to current bounds~\cite{Kohri:2025,Borah:2025}, and a nonzero $\nu_\tau$ flux component at 220~PeV follows from the averaged mixing for the assumed operator. (v) \emph{Theory-side.} The construction requires a gauged discrete flavor symmetry and nonthermal production of the relic; neither is directly observable, and testing them requires embedding the checklist in an explicit completion, which is separate work. (vi) \emph{Cross-check.} The two-insertion reading requires $\Delta L = 2$, so an observation of neutrinoless double-beta decay would establish the Majorana character that branch assumes, while Dirac neutrinos select the one-insertion mediator; proton decay probes the gauge sector above the window and remains complementary. The reading also degrades gracefully: if no further events arrive as the exposure grows, the same analysis converts into the strongest limit on $\nu\bar\nu$-decaying relics at $m_X \approx 4\times10^{8}$~GeV, pushing $\tau/f$ upward and, through $M \propto \tau^{1/6}$, the inferred mediator scale with it.

One event cannot settle the question, but the reading is sharp, the arithmetic definite. The chain is conditional at each link, and its conclusion is concrete: if the 220~PeV feature is a two-body decay line from a cosmologically long-lived flavored onium, and if the leading operator carries one electroweak-Higgs insertion with an $O(1)$ coefficient, then the event points to the Majorana sector of unified models, with a singlet mediator at $M/|C_7|^{1/3} \approx 10^{16} f^{1/6}$~GeV, spanning $10^{14}$--$10^{16}$~GeV across the allowed fraction, on the Dirac branch, or the seesaw window on the $\Delta L = 2$ branch, robust at the order-of-magnitude level. This remains noteworthy without implying evidence for unification: it is the first attempt to read a superheavy scale off a detected neutrino. Each link has an observable consequence, and the first discriminants, morphology and width, arrive with the completed ARCA array; a detector-level line likelihood folded through the KM3NeT response is the natural next step. If the discriminants favor the line, the most energetic neutrino yet detected points to the Majorana-mass scale of grand unification; if they favor the peak, the same event probes black-hole gap acceleration on warm dust.

\begin{acknowledgments}
V.B.~gratefully acknowledges support from the U.S.~Department of Energy, Office of Science, Office of High Energy Physics, under Award Number~DE-SC0017647 and from the William F.~Vilas Trust Estate.
\end{acknowledgments}

\clearpage
\onecolumngrid
\vspace{0.4em}
\begin{center}
\rule{0.5\textwidth}{0.4pt}\\[0.2em]
\textbf{End Matter}
\end{center}
\vspace{0.2em}
\twocolumngrid

\textit{Onium lifetime ladder.} With $|\psi(0)|^2 = (\alpha_{\rm eff} m_X/4)^3/\pi$, $\alpha_{\rm eff} \approx 0.05$, and $m_X = 4.4\times10^{8}$~GeV, Eq.~(\ref{eq:onium}) at $\tau = 10^{30}$~s gives $G_{\rm eff} \approx 1.8\times10^{-47}\,\kappa^{-1/2}$~GeV$^{-2}$, i.e., $2.5\times10^{-46}$~GeV$^{-2}$ at the Coulombic $\kappa = 1/(64\pi)$. Inverting the decomposition $G_{\rm eff} = v_H^{\,d-6}/M^{\,d-4}$, with each range spanning $\kappa = 1/(64\pi)$ to 1,
\begin{center}
\fbox{\begin{minipage}{\dimexpr\columnwidth-2\fboxsep-2\fboxrule}
\vspace{-0.9em}
\begin{align}
M_{d=6} = G_{\rm eff}^{-1/2} &\approx 0.6\text{--}2\times10^{23}~\mathrm{GeV}, \nonumber\\
M_{d=7}^{(v_H=v)} = (v/G_{\rm eff})^{1/3} &\approx 1\text{--}2.4\times10^{16}~\mathrm{GeV}, \nonumber\\
M_{d=8}^{(v_H=v)} = (v^2/G_{\rm eff})^{1/4} &\approx 4\text{--}8\times10^{12}~\mathrm{GeV}, \nonumber\\
M_{d=7}^{(v_H=m_X)} = (m_X/G_{\rm eff})^{1/3} &\approx 1\text{--}3\times10^{18}~\mathrm{GeV}.
\label{eq:ladder}
\end{align}
\vspace{-1.1em}
\end{minipage}}
\end{center}
The first line of Eq.~(\ref{eq:ladder}) is the pure four-fermion operator, trans-Planckian and forbidden; the second is the single electroweak-Higgs insertion, landing at the unification scale; the third is the double insertion, the seesaw window; and the fourth replaces the Higgs by momentum-like insertions. A Planck-suppressed $d=6$ coefficient would give $\tau \sim 10^{12}$--$10^{15}$~s depending on the Planck convention, in every case far short of the age of the universe, which is why the flavor symmetry must be gauged~\cite{Krauss:1989}. Example operators: $\mathcal{O}_7 = (\bar Q' i\gamma_5 Q)(\bar L \tilde H \nu_R)/M^3$ (Dirac, $\bar\nu\nu$) and $\mathcal{O}_8 = (\bar Q' i\gamma_5 Q)(LH)(LH)/M^4$ (Majorana, $\Delta L = 2$, $\nu\nu$), the onium-dressed Weinberg operator~\cite{Weinberg:1979}; the pseudoscalar dark bilinear couples the $^1S_0$ ground state at leading order, while a scalar bilinear projects onto the velocity-suppressed $P$ wave, and the electroweak factor of $\mathcal{O}_7$ is the gauge-invariant Dirac-neutrino Yukawa structure (for systematic bases with right-handed neutrinos see~\cite{Lehman:2014,Bhattacharya:2015,Liao:2017}). The chirality structure shifts $\kappa$ but not the line energy $m_X/2$.

\textit{Event-level line normalization.} The bound-consistent lifetime follows from public numbers, with the detector acceptance cancelling because the line sits at the reference energy. The joint fit across the KM3NeT, IceCube, and Auger exposures quotes a single-flavor $\nu+\bar\nu$ intensity $E^2\Phi \approx 7.5\times10^{-10}$~GeV\,cm$^{-2}$\,s$^{-1}$\,sr$^{-1}$ across the 90\% band~\cite{KM3NeT:landscape}; integrating over the band gives an equivalent intensity $I_1 \approx 1\times10^{-17}$~cm$^{-2}$\,s$^{-1}$\,sr$^{-1}$ at 220~PeV. The Galactic decay line delivers $I = (f/4\pi)(D/m_X\tau)\,\xi$, with $D$ the angle-averaged decay column and $\xi \sim 2/3$ the per-flavor multiplicity after mixing. For the event direction near the anticenter, $D \approx 10^{22}$~GeV\,cm$^{-2}$ (about half the all-sky average), with a comparable extragalactic term~\cite{KM3NeT:hdm}, $I = I_1$ requires $\tau/f \approx (1$--$3)\times10^{29}$~s. This is the bound-consistent normalization by construction, the anchor being the joint flux, and it agrees with the dedicated $\nu\bar\nu$ fits~\cite{Kohri:2025,Borah:2025,Aloisio:2025}. The collaboration's detector-level analysis, which includes the $\nu\bar\nu$ channel and prefers masses above $\sim 100$~PeV consistent with $2E_\nu$, instead returns the lifetime that makes the event typical of KM3NeT's own exposure, $\tau \approx 10^{26}$--$10^{27}$~s, in tension with the $\gamma$-ray and neutrino bounds~\cite{KM3NeT:hdm}; the two normalizations differ by $10^{2}$--$10^{3}$ in $\tau/f$ but a factor $2$--$3$ in the scale. The two spectra are also distinguishable in shape, the line's Galactic edge plus redshifted shoulder against the broad channels' smooth continuum, so accumulated events separate them. Folded through Eq.~(\ref{eq:gut}) this fixes $M/|C_7|^{1/3} \approx (3$--$4)\times10^{15}$~GeV at the fiducial $f = 10^{-2}$ and $(7$--$8)\times10^{15}$~GeV at $f = 1$. At this normalization the single detection is an upward fluctuation of KM3NeT's own exposure, of order the quoted $2.2\sigma$~\cite{KM3NeT:landscape}; that trade is intrinsic to every bound-consistent reading, ours included. The halo fraction of $X$ is taken equal to the cosmological $f$: a cold, collisionless relic with adiabatic perturbations clusters identically to the dominant dark matter, since the collisionless dynamics carries no particle-mass dependence, with free streaming and dynamical friction negligible at this mass; if the dominant component were instead ultralight and cored, $X$ would cluster at least as strongly, making $f D$ conservative. The tracing is fraction-independent, a trace population being a test distribution in the common potential at any $f$: birth momenta redshift below $m_Q$ deep in the radiation era, leaving negligible free streaming; shot noise is irrelevant at these densities; and the floor's $O(1)$ decayed fraction depletes halo and cosmic densities alike, absorbed by defining $f$ today. This is an order-of-magnitude public-area estimate, not a detector-level likelihood: the acceptance cancellation is approximate, since the $E^{-2}$ normalization integrates the full energy and direction response while a line samples one energy and direction, the broad muon-to-neutrino energy likelihood entering both; the $O(1)$ factors in $D$, the extragalactic fraction, $\xi$, and the acceptance recast are the stated uncertainties; a folded likelihood is the necessary next step.

\textit{Relic abundance and binding.} The line flux fixes $f/\tau$; $\Omega_X h^2 = 0.12 f$ corresponds to $n_X/s \approx 10^{-18} f$. Production from the thermal bath during reheating, with Boltzmann-suppressed yield $\propto e^{-2m_Q/T_{\rm RH}}$ up to prefactors set by the production operator, reaches this for $T_{\rm RH} \approx m_Q/(20$--$30)$~\cite{Chung:1998}, with gravitational production a floor~\cite{Garny:2016} and the unitarity bound~\cite{Griest:1990} excluding a thermalized population. Dark confinement below the Bohr momentum binds all survivors at the phase transition; same-flavor onia annihilate, flavored onia at $10^{-18}$ per unit entropy are collisionless thereafter, and binding and survival fractions are $O(1)$ factors absorbed into $f$. The order of magnitude of $f$ is a production question, not a free dial: gravitational production fixes it as a function of $m_Q$ and the inflationary scale~\cite{Garny:2016}, production from the bath as a function of the portal coupling and reheating history~\cite{Chung:1998}, so a specified completion predicts it. The two-component sector is the no-coincidence option: nothing analogous to freeze-out ties this yield to $\Omega_{\rm DM} h^2 = 0.12$, so $f = 1$ would be an accident, whereas the axion misalignment abundance, fixed by the strong-CP solution, naturally saturates the dark matter; the value $10^{-2}$ is representative, not derived. Observation bounds it from below: the flux fixes $f/\tau$, so survival to the present, $\tau \gtrsim t_0$, requires $f \gtrsim t_0(f/\tau) \approx 4\times10^{-12}$, quoted at the conservative end of the band, $\tau/f = 10^{29}$~s (at $10^{30}$~s the floor is $4\times10^{-13}$), and near the floor the decayed fraction reshapes the shoulder and suppresses the edge-to-shoulder ratio, giving spectroscopy direct sensitivity to small $f$. Across the allowed range the sixth root compresses twelve decades of $f$ into two of scale, $M/|C_7|^{1/3} \approx 10^{14}$~GeV at the floor (where $\tau = t_0$ pins it) to $10^{16}$~GeV at $f = 1$, spanning $y \approx 0.4$--$2$ in the seesaw relation for $m_\nu = 0.05$--$0.01$~eV: every allowed $f$ lands the mediator inside the Majorana window, so the branch identification is $f$-robust while the position within the window awaits the production calculation.

\textit{Line shape and flavor propagation.} The predicted spectral shape is shown in Fig.~\ref{fig:lineshape}, whose dashed verticals place the measured counterpart redshifts $z = 0.870$ and $1.227$~\cite{KM3NeT:blazars} at 118 and 99~PeV, on the shoulder rather than at the edge. Under the line hypothesis the counterpart coincidence is therefore a chance alignment and the observed 220~PeV energy reads as decay at $z \approx 0$, dominantly the Galactic column; under the flare hypothesis the same redshifts instead fix the emitted energy near $(1+z)\times220~\mathrm{PeV} \approx 0.4$~EeV at the source. More generally $m_X = 2E_\nu(1+z_{\rm origin})$, so a single event fixes the mass only up to the redshift of its origin; the edge, pinned by the Galactic component once events accumulate, removes the ambiguity. One reading is excluded outright: taking 220~PeV as the observed energy of a line emitted at the counterpart itself, i.e., decays in a dark-matter concentration bound to the blazar with $m_X = 2(1+z)\times220~\mathrm{PeV} \approx 8\times10^{8}$~GeV, fails on column arithmetic, since decay flux is linear in the dark-matter column and even a maximal bound concentration of $\sim 10^{9}\,M_\odot$ at the counterpart distance supplies of order $10^{-14}$ of the Milky Way halo column; the lifetime that yields one blazar-line event would flood the detectors with halo events. The edge is therefore calibrated locally, at $z \approx 0$. One prior does enter the energy itself: the 220~PeV median unfolds the reconstructed muon energy under an $E^{-2}$ spectral assumption~\cite{KM3NeT:Nature}; a monochromatic prior reweights the same muon-energy likelihood and shifts the median somewhat, a dependence bracketed by the 90\% interval propagated into the $m_X$ band throughout. Electroweak final-state radiation dresses $X \to \nu\bar\nu$ at $m_X \sim 4\times10^{8}$~GeV with a soft continuum carrying a modest energy fraction, treated and constrained in Refs.~\cite{Kohri:2025,Borah:2025}; the hard spike remains resolution-narrow. For halo and intergalactic propagation the oscillation phases decohere and the matter potential $\sqrt2 G_F n_e$ stays many orders of magnitude below $\Delta m^2/2E$ (the MSW-resonant density at 220~PeV is $\sim 10^{-7}$~g\,cm$^{-3}$, approached only deep inside accretion-flow environments irrelevant to halo decay), so decohered vacuum mixing applies: the Earth ratio is $\sum_i |U_{\alpha i}|^2 |U_{\beta i}|^2$ acting on the source flavor composition; every flavor exceeds $\approx 15\%$ for a democratic or incoherent source, e.g., $(1:0:0)_S \to (0.55:0.17:0.28)_\oplus$ and $(0:1:0)_S \to (0.17:0.45:0.38)_\oplus$; a single source mass eigenstate is the exception (pure $\nu_3$ gives an electron fraction of a few percent).

\textit{Warm-dust peak backbone.} The gap potential $e\Phi \approx 2\times10^{20}\,\mathrm{eV}\,(a/M)(B/10^4\,\mathrm{G})(M/10^9 M_\odot)$~\cite{Boldt:1999} with gap fraction $\eta \approx 0.02$ places the required 4.4~EeV a factor $\sim 2$ below the curvature-radiation ceiling, where operation sharpens the beam. The $\Delta$-resonance condition on a thermal field~\cite{Stecker:1968,Mucke:2000,Huemmer:2010} selects $\varepsilon_0 \approx 0.036$--$0.07$~eV (head-on to $90^\circ$), dust at $T \approx 150$--$300$~K, and a photopion optical depth $\tau_{p\gamma} \approx n_{\rm IR}\sigma_\Delta R \approx 8\,L_{45}(R/\mathrm{pc})^{-1}(\varepsilon_0/0.036~\mathrm{eV})^{-1}$ reaching order unity after thermal, angular, and geometric dilution. Muon damping needs $B \gtrsim 10$~G at $E_\mu \approx 0.7$~EeV, deep in the torus cavity, so the composition is generically mixed. Parent protons never arrive: neutrons ($\approx 3.5$~EeV) decay within $\sim 32$~kpc, protons deflect and arrive $> 10^4$~yr late, and the $\pi^0$ cascade falls below the Fermi-LAT bound~\cite{Fang:2025}.

\end{document}